\documentclass[twocolumn,showkeys,preprintnumbers,longbibliography,amsmath,amssymb,epsfig,floatfix,prx]{revtex4-2}
\usepackage[utf8]{inputenc}
\usepackage{ulem}
\usepackage{physics}
\usepackage{amsmath}
\usepackage{mathtools}
\usepackage{xcolor}
\usepackage{color, colortbl}
\definecolor{LightCyan}{rgb}{0.88,1,1}
\usepackage{physics}
\DeclareMathAlphabet{\altmathcal}{OMS}{cmsy}{m}{n}
\usepackage{siunitx}
\usepackage{multirow}
\usepackage{amsfonts}
\usepackage{amssymb}
\usepackage{bm}
\usepackage{calrsfs}
\DeclareMathAlphabet{\mathcalligra}{T1}{calligra}{l}{m}
\usepackage{epsfig}
\usepackage{graphicx}
\usepackage{dcolumn}
\usepackage{color}
\usepackage{natbib}  
\usepackage{hyperref}
\usepackage{breakurl}
\hypersetup{colorlinks=true, citecolor=blue, urlcolor=blue, linkcolor=blue}
\usepackage{bm}
\usepackage{booktabs}
\usepackage[version=3]{mhchem}
\usepackage[english]{babel}

\newcolumntype{L}[1]{>{\raggedright\arraybackslash}p{#1}}
\newcolumntype{C}[1]{>{\centering\arraybackslash}p{#1}}
\newcolumntype{R}[1]{>{\raggedleft\arraybackslash}p{#1}}

\definecolor{green}{RGB}{0,100,0}
\definecolor{violet}{RGB}{142, 68, 173}
\definecolor{edit}{RGB}{0, 128, 255}

\begin{document}

\title{Emergent magnetic and charge ordered phases in freestanding ultrathin \ce{LaVO3}}
\author{Ashutosh Anand}
\author{Mukul Kabir}
\email{mukul.kabir@iiserpune.ac.in}
\affiliation{Department of Physics, Indian Institute of Science Education and Research, Pune 411008, India}


\begin{abstract} 
	Transition metal oxide perovskites are an ideal platform for exploring the interplay between spin, orbital, charge and lattice degrees of freedom. Among them, \ce{LaVO3} has been extensively studied in heterostructures and superlattices, where exotic phases have been reported. Motivated by the advances in freestanding oxide membranes, we investigate the intrinsic properties of freestanding ultrathin \ce{LaVO3} films using density functional theory. Our calculations reveal a sequence of magnetic phase transitions with thickness, starting from stripe-AFM in monolayer until the bulk like C-AFM is recovered. Beyond four layers, polar catastrophe driven charge transfer dopes the surface layers giving rise to stripe-AFM and ferromagnetic surface states while the central layers remain bulk like. We further explore this fact by studying charge doped monolayer, discovering that hole doping drives the system into ferromagnetic state. Doping also induced charge ordering in the system. A striped charge ordering pattern is observed at 0.5 h/fu, while a 3:1 stripe pattern emerges at 0.25 h/fu, indicating that the periodicity of the superstructure changes with doping concentration.
\end{abstract} 

\maketitle
\section{Introduction} 
The discovery of graphene ignited broad interest in two-dimensional (2D) materials as platforms for quantum phenomena and tunable functionalities. Transition metal oxides (TMOs) occupy a particularly compelling position in this landscape, governed by the interplay of strong electron correlations, spin-orbit coupling, and lattice degrees of freedom. Decades of research have uncovered a remarkable diversity of many-body phases, including high-temperature superconductivity in cuprates, colossal magnetoresistance in manganites, multiferroic order, and interfacial two-dimensional electron gases (2DEGs)~\citep{bednorz1986possible, jin1994thousandfold, fiebig2016evolution, ohtomo2004high}. 

Reducing TMOs to the ultrathin limit provides access to regimes where quantum confinement and symmetry breaking can stabilize or suppress these phases in ways inaccessible in bulk. Conventionally, however, ultrathin TMO films must be grown epitaxially on a structurally compatible substrate, which restricts the accessible material space and introduces epitaxial clamping that can mask or alter intrinsic electronic and magnetic properties. Recent advances in fabricating freestanding perovskite oxide membranes overcome this constraint~\citep{lu2016synthesis, chiabrera2022freestanding}. Striking examples include freestanding \ce{LaMnO3} films at 4~nm exhibiting soft ferromagnetism~\citep{xiang2024boosting}, \ce{La_{0.7}Ca_{0.3}MnO3} membranes undergoing a metal-insulator transition under uniaxial and biaxial strain~\citep{hong2020extreme}, \ce{BiFeO3} films below 3 unit cells transitioning to a tetragonal phase with enhanced polarization~\citep{ji2019freestanding}, and \ce{SrTiO3} membranes displaying strain-induced room-temperature ferroelectricity~\citep{xu2020strain}. Together, these results establish freestanding perovskite oxide membranes as a powerful new platform for uncovering emergent phases at the ultrathin limit.

Electronic instability arising from degenerate, partially filled $B$-site $d$-orbitals drives Jahn-Teller (JT) distortion in $ABO_{3}$ perovskites, causing individual $BO_{6}$ octahedra to spontaneously break their local cubic symmetry. This local deformation propagates cooperatively through the lattice, coupling with steric instabilities imposed by ionic size mismatch. When the $A$-site cation is small relative to the $BO_6$ framework, quantified by a Goldschmidt tolerance factor $t < 1$, the rigid octahedral network undergoes collective tilting and rotation to optimize packing, typically stabilizing a \ce{GdFeO3}-type orthorhombic structure. The resulting lattice distortions couple with spin, orbital, and charge degrees of freedom, lifting quantum degeneracies and driving the emergence of novel electronic phases.

This interplay is particularly striking in the $t^2_{2g}$-perovskites \ce{LaVO3} and \ce{YVO3}, both containing \ce{V^{3+}}  ions~\citep{bordet1993structural,mahajan1992magnetic,PhysRevB.52.324,miyasaka2003,fang2004quantum,tung2008,kawano1994magnetic, PhysRevLett.87.245501,PhysRevB.65.174112,PhysRevLett.91.257202,PhysRevLett.101.245702,PhysRevB.69.054111, PhysRevB.105.094412}. The twofold orbital degeneracy of the $t^2_{2g}$ manifold drives a cooperative JT effect, triggering long-range orbital ordering.  Orthorhombic ($Pbnm$) \ce{LaVO3} remains paramagnetic and orbitally ordered above 140~K. Below 140~K, cooperative JT distortion simultaneously lowers the crystal symmetry to monoclinic ($P2_1/a$) and stabilizes G-type orbital ordering and C-type AFM ordering, which persist to the lowest temperatures~\citep{bordet1993structural,mahajan1992magnetic,PhysRevB.52.324,miyasaka2003,fang2004quantum,tung2008}. The relatively large \ce{La^{3+}} cation maintains a sufficiently symmetric framework that a single orbital configuration remains energetically dominant, precluding any further structural or magnetic reorganization upon cooling. In contrast, \ce{YVO3} displays a richer, temperature-driven sequence of structural and magnetic phase transitions, driven by the stronger orthorhombic distortion imposed by the smaller \ce{Y^{3+}} cation, which frustrates the orbital and magnetic degrees of freedom more severely. A structural transition at 200~K simultaneously lowers the crystal symmetry to $P2_1/a$ and stabilizes G-type orbital ordering~\citep{kawano1994magnetic, PhysRevLett.87.245501,PhysRevB.65.174112,PhysRevLett.91.257202,PhysRevLett.101.245702}. Below 116~K, C-type AFM ordering develops without further change in crystal symmetry. Below 77~K, the progressive increase in octahedral tilting destabilizes the C-type orbital and magnetic order, driving a first-order transition back to the $Pbnm$ symmetry, accompanied by C-type JT distortion and G-type AFM ordering. Phonon anomalies and spin-lattice coupling across this transition have been further characterized by neutron scattering and infrared spectroscopy~\citep{PhysRevB.69.054111, PhysRevB.105.094412}. This temperature-driven reversal of both orbital and magnetic order is absent in \ce{LaVO3}, which has a more rigid orbital landscape, underscoring the exceptional sensitivity of \ce{YVO3} to competing electron-lattice interactions. In \ce{LaVO3}, the narrow $t_{2g}$ bandwidth stabilizes a strongly correlated Mott-Hubbard state, where competing electronic correlations and lattice distortions become particularly sensitive to spatial confinement and structural relaxation in freestanding ultrathin films.

Interest in \ce{LaVO3} gained momentum following the discovery of a polar catastrophe driven 2DEG at the \ce{LaVO3}/\ce{SrTiO3}~\citep{hotta2007polar} and \ce{LaVO3}/\ce{KTaO3}~\citep{wadehra2020planar} interfaces. These interfaces also host unconventional superconductivity at low temperatures~\cite{halder2022unconventional, liu2023superconductivity}, absent in either bulk constituent, as well as the planar Hall effect and anisotropic magnetoresistance~\citep{wadehra2020planar, tomar2021anisotropic}. Beyond heterointerfaces, \ce{LaVO3}/\ce{SrVO3} superlattices display a metal-insulator transition accompanied by room-temperature ferromagnetism~\citep{sheets2007effect, luders2009room}. Collectively, these observations establish \ce{LaVO3}-based heterostructures as a versatile platform for exotic physical phenomena, yet the microscopic origin of many of these effects remains unresolved. A key missing ingredient is a clear understanding of the intrinsic properties of \ce{LaVO3}, decoupled from substrate strain and interfacial charge transfer. The freestanding membrane geometry eliminates these extrinsic effects. Here, we investigate the electronic and magnetic properties of freestanding ultrathin \ce{LaVO3} films from first principles, providing a theoretical baseline for understanding the emergent behavior of \ce{LaVO3}-based heterostructures and superlattices.

We uncover a rich, thickness driven evolution of magnetic order in freestanding \ce{LaVO3}, marked by a sequence of distinct magnetic transitions as the film is reduced to the monolayer limit. In thicker films, a polar catastrophe induced electronic reconstruction at the surface stabilizes emergent in-plane ferromagnetic and stripe-ordered phases that coexist with the bulk-like C-type antiferromagnetic order of the interior. These surface magnetic phases are accompanied by charge ordering, which drives an insulator-to-insulator transition and emerges as the dominant mechanism governing the electronic ground state in ultrathin \ce{LaVO3}.

\section{Methodology}
First-principles calculations are performed within the framework of density functional theory (DFT)~\citep{PhysRev.136.B864, PhysRev.140.A1133}, as implemented in the Vienna \textit{ab initio} Simulation Package~\citep{PhysRevB.48.13115, PhysRevB.54.11169}. The electronic wave functions are expanded using the projector-augmented wave formalism~\citep{PhysRevB.50.17953}, with the kinetic energy cutoff for the plane-wave basis set to 500 eV. The exchange-correlation energy is described using the Perdew Burke-Ernzerhof functional~\cite{PhysRevLett.77.3865}, To account for the strong electron correlation among the vanadium $3d$ states, we supplement the functional with a Hubbard-like on-site Coulomb repulsion $U$, using the rotationally invariant scheme~\citep{PhysRevB.57.1505}. An effective $U$ of 3 eV is considered, consistent with prior studies on vanadates~\citep{fang2004quantum, varignon2015coupling, Dai_2018}. Spin-orbit coupling (SOC) is included in all calculations unless otherwise stated. 

\begin{figure}
		\includegraphics[width=0.6\linewidth,angle=90]{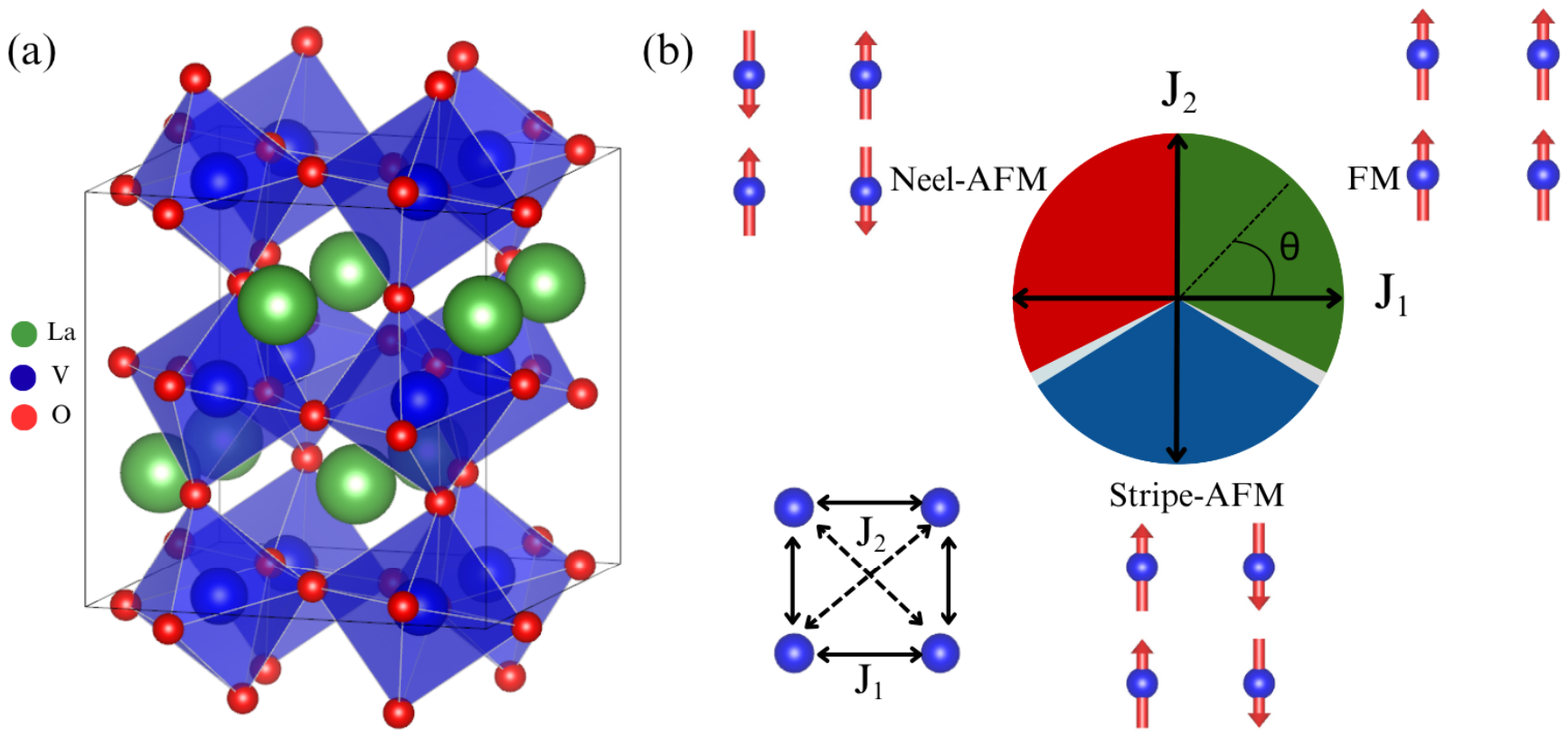}
		\caption{(a) $\sqrt{2}\cross\sqrt{2}$ supercell of the orthorhombic structure. (b) Phase diagram of the $J_1-J_2$ Heisenberg model, depending on the frustration angle $\theta = \tan^{-1}(J_2/J_1)$ the ground state can be FM, Neel-AFM or stripe-AFM. The gray region in the phase diagram corresponds to the spin-liquid regime. }
		\label{fig1}
	\end{figure}

For bulk $\text{LaVO}_3$, we adopted the orthorhombic perovskite structure as the starting point. A $\sqrt{2} \times \sqrt{2} \times 1$ supercell [Fig. \hyperref[fig1]{1(a)}] was constructed to accommodate the stripe-AFM configuration. Thin films of varying thickness were modeled using this in-plane supercell $\sqrt{2} \times \sqrt{2}$ geometry, with a vacuum gap of at least 15 \AA\ added to suppress spurious interactions between periodic images along the out-of-plane direction. The Brillouin zone is sampled using $\Gamma$-centered Monkhorst-Pack $k$-mesh of $9\times9\times9$ for bulk and  $9\times9\times1$ for thin-film geometries. Structural relaxation is carried out until the Hellmann-Feynman forces on all atoms are converged below 0.01 eV/\AA. For the freestanding thin-film calculations, both the in-plane lattice parameters and all atomic positions are fully relaxed to simulate the freestanding condition.

The magnetic interactions in \ce{LaVO3} are described by a Heisenberg model on a square lattice of magnetic \ce{V^{3+}} ions, accounting for first- and second-neighbour exchange interactions. The spin Hamiltonian, including single-ion anisotropy, takes the form,
\begin{equation}
\altmathcal{H} = -\sum_{i < j} J_{ij} \mathbf{S}_i \cdot \mathbf{S}_j - \sum_i \left[ A_z (S_i^z)^2 + A_{xy} (\Delta S_i^2) \right], \nonumber
\end{equation}
where $J_{ij}$ denotes the isotropic exchange coupling between sites $i$ and $j$, encompassing the first- and second-neighbour in-plane interactions $J_1$ and $J_2$, and the interplane coupling $J_\perp$. A positive (negative) value of $J_{ij}$ corresponds to FM (AFM) exchange between $\bm{S_i}$ and $\bm{S_j}$. $A_z$ accounts for the uniaxial single-ion anisotropy, where $A_z > 0$ ($A_z < 0$) denotes easy-axis (easy-plane) magnetism. $A_{xy}$ captures the in-plane anisotropy that breaks the continious rotational symmetry within the $xy$-plane, and selects a preferred spin orientations within the plane, and $\Delta S_i^2 = (S_i^x)^2 - (S_i^y)^2$. The in-plane magnetic order is governed by the ratio $J_2/J_1$, which stabilizes N\'eel-AFM, stripe-AFM, or FM phases depending on the exchange regime [Fig. \hyperref[fig1]{1(b)}]. These in-plane configurations, in conjunction with the interplane coupling $J_\perp$, determine the overall three-dimensional magnetic ground state. We compute the isotropic exchange interactions through energy mapping of various spin-ordered phases. To investigate the magnetic phase transitions and determine the ordering temperature, we performed classical Monte Carlo simulations using the Vampire code~\citep{Evans_2014}.
 

\section{Results and Discussion}
We first characterize bulk \ce{LaVO3}, where the close agreement between the present results and prior experimental and theoretical studies serves as a benchmark. Building on this foundation, we investigate the evolution of the electronic and magnetic properties in freestanding ultrathin films.

\subsection{Electronic and magnetic structure of bulk \ce{LaVO3}}

Bulk \ce{LaVO3} crystallizes in the \ce{GdFeO3}-type orthorhombic $Pbnm$ structure at room temperature, with cooperative \ce{VO6} octahedral tilts and rotations ($a^-a^-c^+$) driven by the ionic size mismatch of \ce{La^{3+}} ($t < 1$). As the temperature is lowered, spin ordering precedes the structural and orbital ordering. Below 143~K, C-type AFM order develops, with FM alignment along the $c$-axis and AFM coupling in the $ab$-plane. 
 In agreement with experiment, we find that the $Pbnm$ structure stabilizes in C-AFM order. The extracted exchange parameters, $(J_1, J_2, J_\perp)$ = ($-5.11, -1.27, 3.75$) meV are consistent with previous reports. 
 
\begin{figure}
\includegraphics[width=\linewidth]{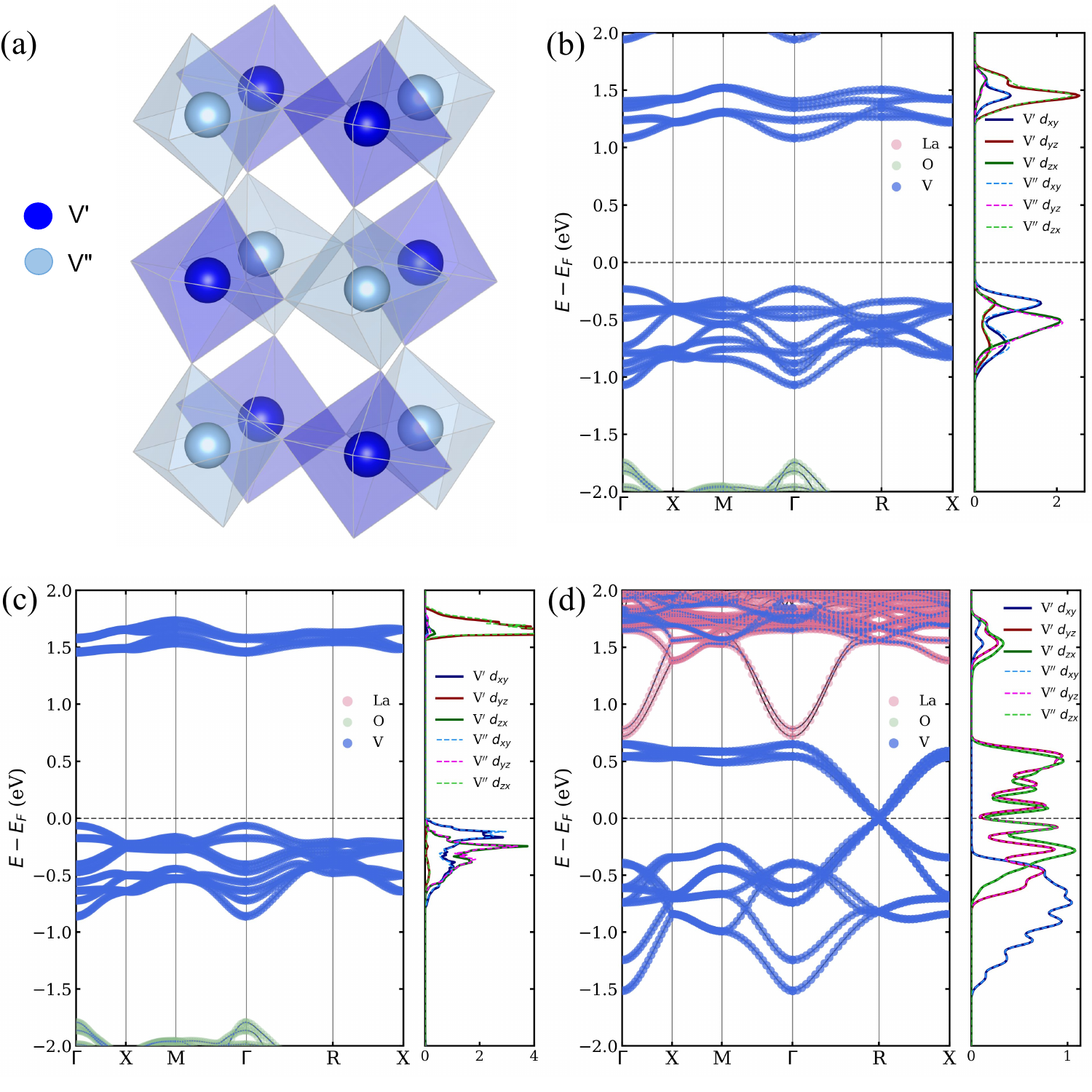}
\caption{(a) Representative crystal structure to illustrate spatial arrangement of V$^{\prime}$ and V$^{\prime\prime}$ atoms. Electronic structure evolution of C-AFM state across different lattice symetries (b) orthorhombic, (c) monoclinic and (d) cubic. Orthorhombic and monoclinic phase exhibits an insulating gap with pronounced orbital ordering, whereas cubic phase demonstrates a metallic nature and no orbital ordering. }
\label{bulkband}	
\end{figure}

Upon further cooling at 140~K, a first-order structural transition lowers the symmetry to the monoclinic $P2_1/b$ phase, and the optimized lattice parameters are in good agreement with the experimental data measured at 10~K. This structural transition simultaneously stabilizes G-type orbital ordering due to the cooperative elongation of adjacent \ce{VO6} octahedra along orthogonal axes. The G-type orbital ordering is characterized by uniform occupation of the $d_{xy}$ orbital at every \ce{V^{3+}} site, with the second electron alternating between $d_{yz}$ and $d_{zx}$ in a fully staggered pattern along all three directions. Within the Kugel-Khomskii superexchange framework, this localized orbital arrangement suppresses hopping along the $c$-axis, favoring FM alignment, while promoting AFM coupling in the $ab$-plane, directly stabilizing the observed C-type AFM order.  Orbital occupancy analysis indicates that while both $Pbnm$ and $P2_1/b$ structures exhibit orbital ordering, it is noticeably enhanced in the $P2_1/b$ symmetry, which concurrently displays a more robust spin ordering (Figure~\ref{bulkband}).  

The calculated spin moment of $1.8~\mu_{\rm B}$ exceeds the experimentally measured value of $\sim 1.3~\mu_{\rm B}$, a discrepancy driven by quantum orbital fluctuations. The incomplete lifting of orbital degeneracy by the weak $t_{2g}$ JT effect allows residual, site-dependent mixing of $d_{yz}$ and $d_{zx}$ states, with calculated occupancies alternating between $(n_{yz}, n_{zx}) \sim (0.8, 0.2)$ and $(0.2, 0.8)$ in the $Pbnm$ phase. Competing with static structural distortions, these quantum orbital fluctuations systematically suppress the ordered magnetic moment and lower the observed transition temperature. This scenario alters dramatically in the low-temperature $P2_1/b$ structure,  where orbital polarization becomes pure and orbital mixing is negligible. 

 \ce{LaVO3} is a Mott insulator within both the $Pbnm$ and $P2_1/b$ symmetries, featuring a calculated electronic gap that ranges between 1.4 and 1.5~eV, in close agreement with the experimental optical gap (Figure~\ref{bulkband}). An analysis of the valence and conduction bands indicates weak $p\text{-}d$ hybridization, while the relatively narrow $t_{2g}$ bandwidth (0.8 - 0.9 eV) underscores strong electronic correlations in the system.

	\begin{figure*}
	\includegraphics[width=\linewidth]{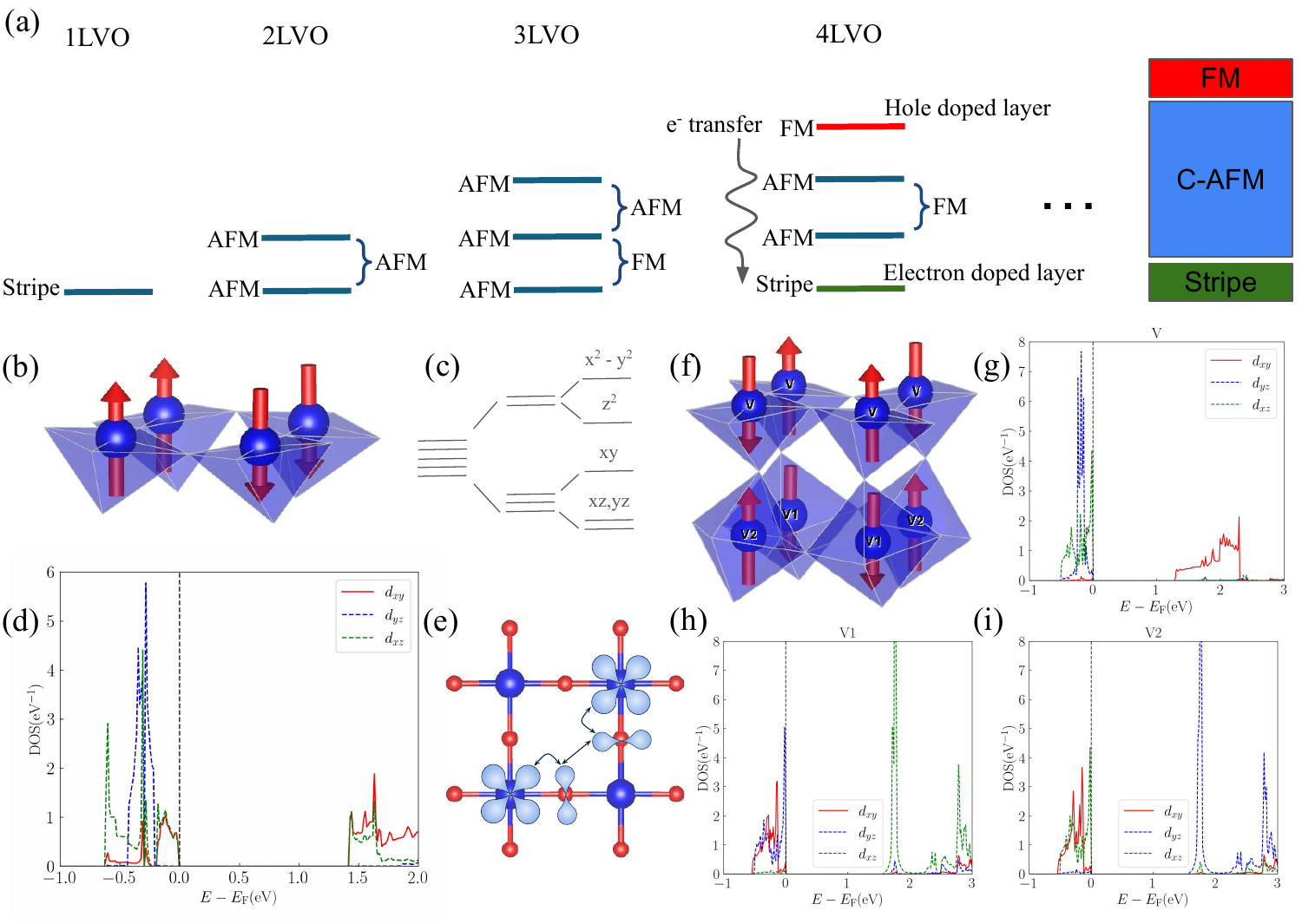}
	\caption{(a) Schematic evolution of magnetic ground state in \ce{LaVO3} thin films as a function of thickness. (b) Structure of 1LVO with stripe-AFM configuration. (c) d-level splitting under a square pyramidal crystal field environment. (d) Projected DOS of V $t_{2g}$ manifold showing dominant $d_{xz}$ and $d_{yz}$ orbitals as expected, but also a finite $d_{xy}$ contribution. (e) Illsutration of $d_{xy}$ mediated next neighbor exchange pathway. (f) Structure of 2LVO along with projected DOS of V ions. (g) Vanadium ions on top layer show no orbital ordering and (h),(i) ions at the bottom layer show orbital ordering same as bulk. }
	\label{fig2}
    \end{figure*}

\subsection{Ultrathin \ce{LaVO3} layers}

	Starting from monolayer, our calculations reveal that monolayer \ce{LaVO3} favors stripe-AFM, in contrast to bulk, where in-plane configuration corresponds to Neel-AFM configuration. One of the major differences we observe for monolayer, is that the vanadium atom is no longer octahedrally coordinated. Instead it resides in a five fold coordinated square pyramid environment [Fig. \hyperref[fig2]{3(b)}], which breaks $t_{2g}$ degeneracy and stabilizes $d_{xz}$ and $d_{yz}$ as the lowest energy states [Fig. \hyperref[fig2]{3(c)}], this is what we see in the orbital projected DOS [Fig. \hyperref[fig2]{3(d)}]. Thus, the orbital ordering disappears in monolayer. However, we do see a $d_{xy}$ occupation near the fermi level. This $d_{xy}$ orbital is responsible for a significant next nearest neighbor interaction [Fig. \hyperref[fig2]{3(e)}]. This interaction is further supported by the relatively flat monolayer geometry. The bond angles are near 170$^{\circ}$, which makes O-O hopping significant, compared to bulk where the bond angles are near 150$^{\circ}$. FM state is about 34meV/fu higher and AFM state is about 31meV/fu higher than the stripe ground state this gives us the $J_1$ and $J_2$ as -0.75 meV and -8.12meV respectively. The large ratio of $J_2/J_1$ puts the system in the stripe-AFM regime of phase diagram. 

	As we increase the thickness to bilayer [Fig. \hyperref[fig2]{3(f)}],  the vanadium atoms in the top layer are fivefold coordinated, same as the monolayer case. As a result, orbital ordering is absent and both $d_{yz}$ and $d_{xz}$ orbitals are present at each site [Fig. \hyperref[fig2]{3(g)}]. However, unlike the monolayer, there is no $d_{xy}$ occupation, which kills the next-nearest-neighbor superexchange pathway and drives the in-plane magnetic configuration toward Neel-AFM. In contrast, the bottom layer remains octahedrally coordinated and retains the same orbital ordering [Fig. \hyperref[fig2]{3(h), (i)}] as in the bulk, resulting in the same Neel-AFM in-plane order. Although the $d_{yz}$ and $d_{xz}$ orbitals alternate in the bottom layer, the presence of both orbitals at every site in the top layer leads to substantial interlayer orbital overlap, stabilizing antiferromagnetic interlayer coupling, in contrast to the ferromagnetic interlayer coupling observed in the bulk. When we go to trilayer, the same physics follows. The only difference being that the top two layers behave similar to the bilayer system and are coupled antiferromagnetically, whereas the bottom two layers resemble the bulk environment and are coupled ferromagnetically.

\begin{figure}
		\centering.
		\includegraphics[width=\linewidth]{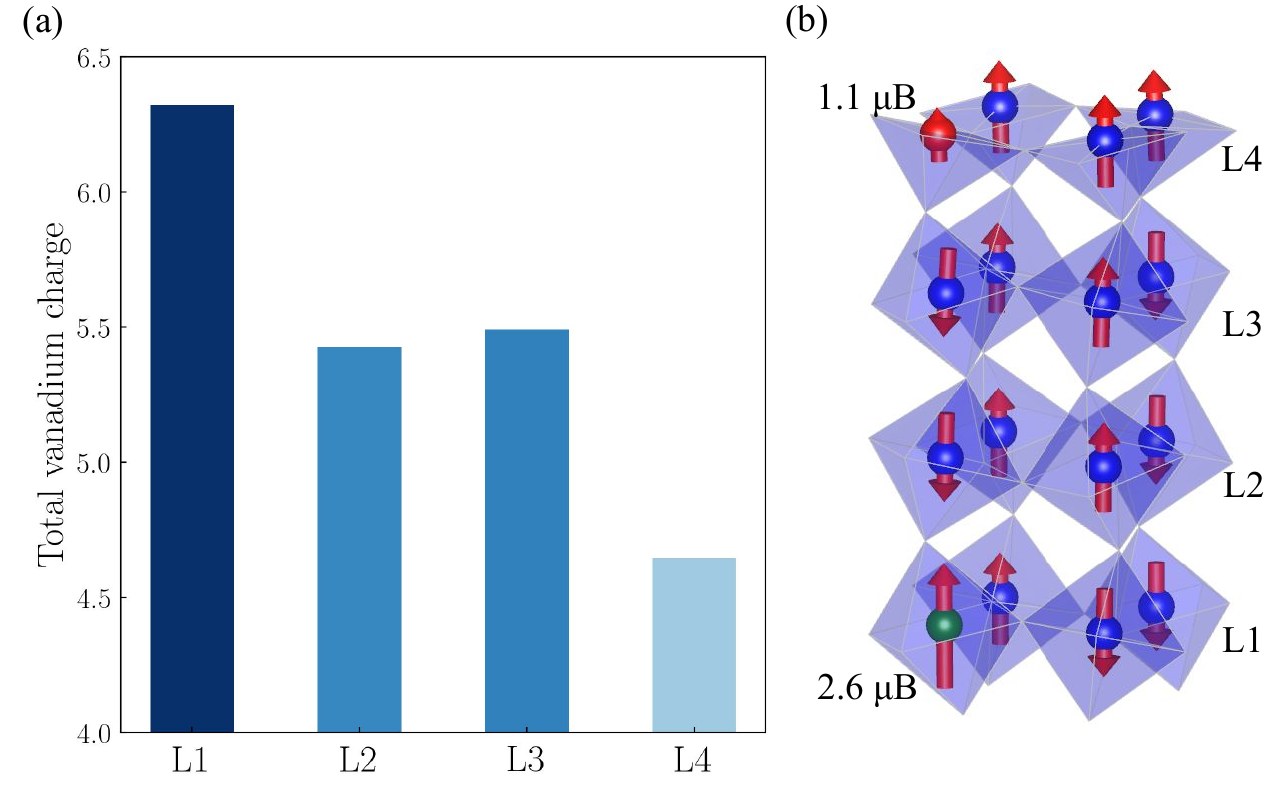}
		\caption{(a) Layer resolved total vanadium charge across 4LVO layers, showing a clear charge redistribution between layers. (b) Structure of 4LVO, showing site selective nature of charge transfer, red and green sphere mark the sites that have lost and gained electrons.}
		\label{fig3}
	\end{figure}

	When the thickness is further increased to four layer or more, a charge transfer is induced [Fig \hyperref[fig3]{4(a)}] due to polar catastrophe, this is consistent with the observation that 2DEG formation at the LVO/STO heterostructure requires 4 layers of LVO \cite{hotta2007polar}. In our case this charge transfer leaves the top layer hole doped and bottom layer becomes electron doped. This doping changes the magnetic configuration of the respective layers. Hole doped layer becomes FM and electron doped layer goes back to stripe-AFM configuration, we discuss this in detail in the following section. To verify the thickness dependence, we extend our calculation to eight layers of LVO. The results confirm that the central, bulk-like layer recover the C-type AFM ground state characteristic of bulk LVO, while the top surface becomes FM and bottom surface becomes stripe-AFM. We also observe that the charge transfer process is site selective, only certain vanadium atoms in hole doped layer seem to have lost electrons and only certain atoms in electron doped layer have gained electrons [Fig \hyperref[fig3]{4(b)}]. This asymmetry indicates onset of charge ordering in doped layers. Similar charge ordering has also been reported in LVO/SVO superlattices where the \ce{VO2} layer is effectively hole doped \cite{park2007chargeorder}. To further understand the microscopic origin of this charge ordering, we look at hole doped monolayer LVO and electron doped bulk LVO. We make this distinction because top layer being five fold coordinate behaves more like monolayer LVO and bottom layer being six fold coordinated behaves more like bulk LVO. 
	
	\subsection{Hole doped monolayer \ce{LaVO3}}
\begin{figure}
			\centering
			\includegraphics[width=\linewidth]{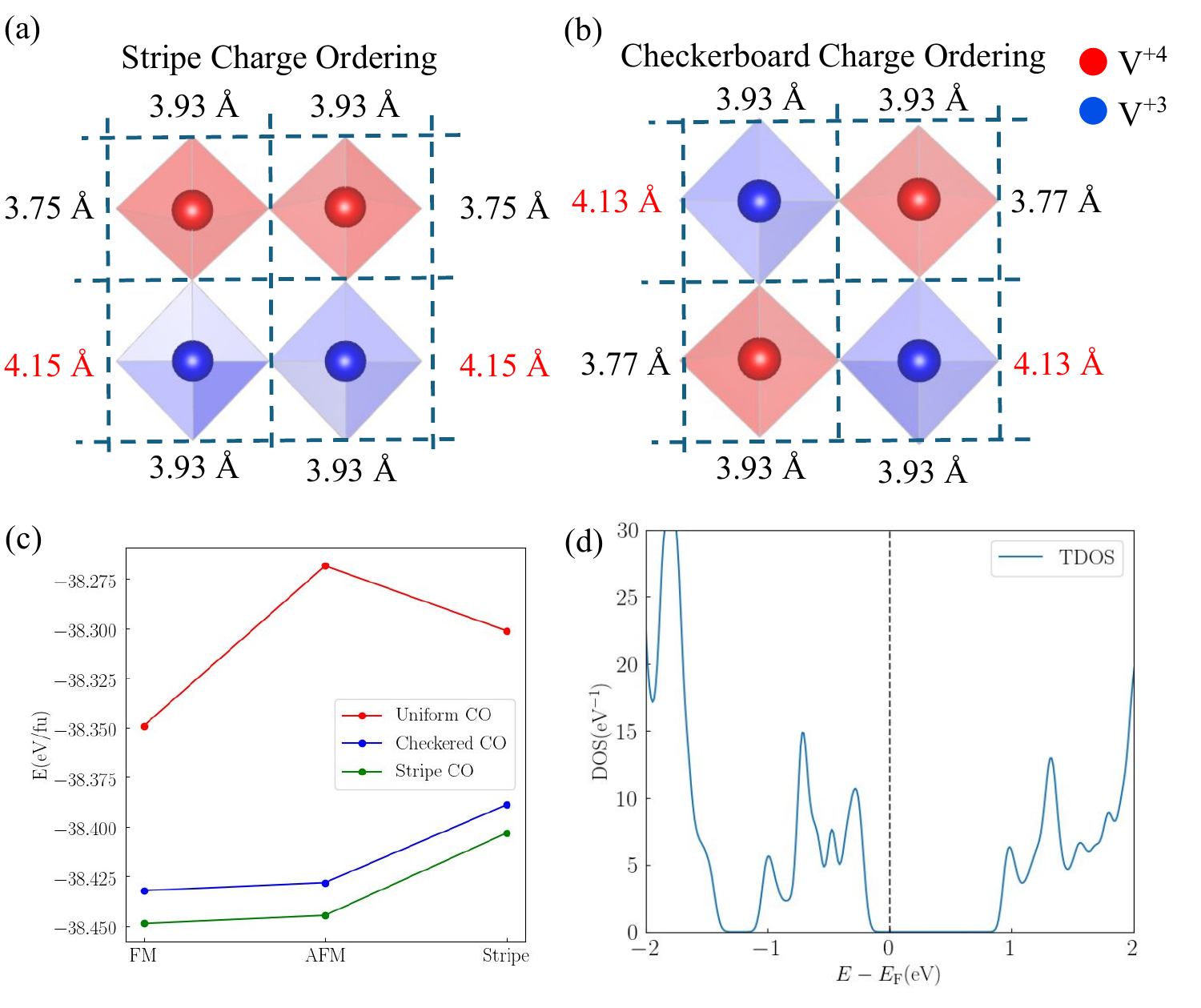}
			\caption{(a) Stripe and (b) Checkerboard charge ordering patterns at a doping concentration of 0.5h/fu. Sites with higher electron density have expanded octahedras while electron depleted sites have contracted octahedras, reflecting breathing mode distortions. (c) Relative energies of charge ordered states under different magnetic configuration. FM state in stripe charge ordering is energetically most favorable. (d)$t_{2g}$ projected DOS for V$^{+3}$ and V$^{+4}$. }
			\label{fig4}
		\end{figure}

		Charge doping directly modifies the orbital occupancy near the Fermi level and is expected to influence magnetic exchange interactions. It is therefore important to understand how the doped carriers are distributed in the system. When the number of electrons per site becomes non-integer, two broad scenarios may arise. In the first case, the doped carriers remain itinerant and hop rapidly between ions, rendering all sites equivalent and resulting in metallic conductivity. In the second case, the carriers become localized, leading to a periodic modulation of the charge density and the formation of a charge-ordered superstructure. Transition metals which can exist in multiple valence states often exhibit such charge ordering phenomena. Hole-doped manganites such as \ce{La$_{1-x}$Ca$_x$MnO3} provide a well-known example of this behavior, at half doping ($x = 0.5$), Mn$^{3+}$ and Mn$^{4+}$ ions order in a checkerboard pattern. Checkerboard ordering is a very natural ordering, which helps in minimizing the Coulomb repulsion. However, Coulomb interaction is not the sole driving force determining the type of ordering, other degrees of freedom such as orbital and lattice might also play a role. We see this in hole doped monolayer LVO, at a hole concentration of 0.5h/fu where V$^{3+}$ and V$^{4+}$ are present in a 1:1 ratio, a stripe type charge ordering emerges as the lowest energy configuration across all magnetic arrangements considered [Fig \hyperref[fig4]{5(c)}]. Among these the FM state is found to be most energetically favourable. In the ground state V$^{3+}$ and V$^{4+}$ have a magnetic moment of 1.95$\rm{\mu_B}$ and 1.14$\rm{\mu_B}$ respectively, giving a net moment of approximately 1.5$\rm{\mu_B}$/fu. With a significant ferromagnetic next nearest neighbor exchange of 11.1meV, the FM state has a transition temperature of 110K.
		
		The charge ordering is accompanied by breathing-mode distortions, characterized by alternating expansion and contraction of the octahedra around V$^{3+}$ and V$^{4+}$ sites [Fig \hyperref[fig4]{5(a), (b)}]. These lattice distortions, together with strong electronic correlations, localize the excess carriers on inequivalent sites, opening a gap at the Fermi level and making  the system insulating [Fig \hyperref[fig4]{5(d)}]. In contrast, the unrelaxed structure does not stabilize charge ordering, indicating that electronic correlations alone are insufficient to localize the carriers, and the system consequently remains metallic.

\begin{figure}
			\centering
			\includegraphics[width=\linewidth]{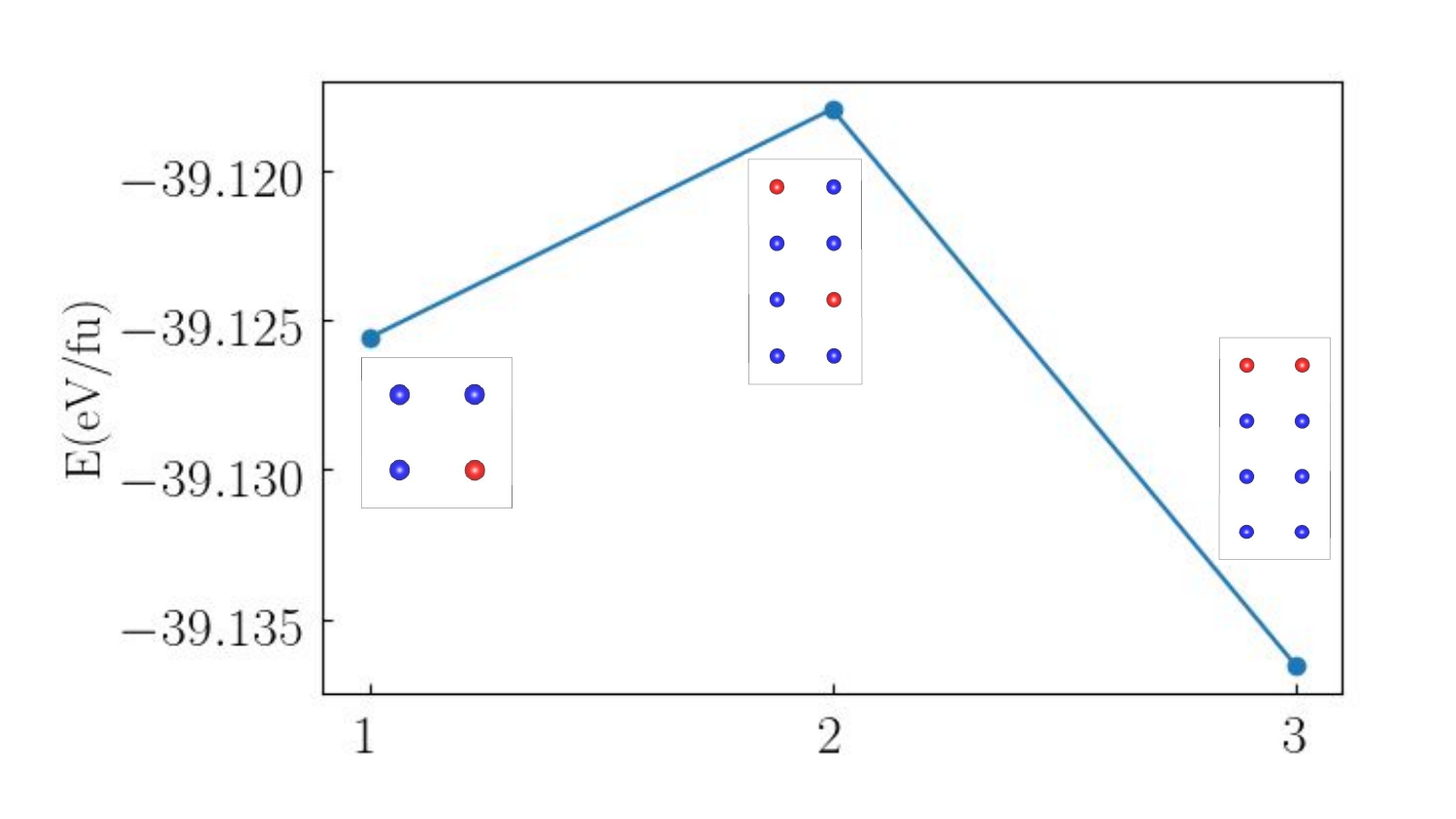}
			\caption{Relative energies of different charge ordering pattern considered at a doping concentration of 0.25h/fu in the FM state. }
			\label{fig5}
		\end{figure}

		Discussing charge ordering at doping levels other than 0.5h/fu is less straightforward. For example, at 0.25h/fu, the charge ordered state corresponds to a 3:1 ratio of V$^{3+}$ and V$^{4+}$ ions. Within the unit cell considered here, only one such configuration is possible. However, considering a larger supercell allows multiple distinct arrangements, and the number of possible charge ordering patterns increases rapidly with system size, making an exhaustive search impractical. Among the configurations we examined, a 3:1 stripe pattern emerges as the lowest energy state [Fig \hyperref[fig5]{6}]. This is physically intuitive as a 1:1 stripe ordering is favored at 0.5h/fu, reducing the doping effectively increases the wavelength of the charge order. This clearly points out at periodic modulation with respect to doping, something that has been observed in doped manganites. The periodicity of the observed superstructure in \ce{La$_{1-x}$Ca$_x$MnO3} for $x > 0.5$ changes with doping \cite{}.
		
		Electron doping the monolayer does not stabilize a charge ordering pattern and the structure becomes metallic. This is probably because charge ordering would require few of the vanadium to be in V$^{2+}$ state which will have a  $t^3_{2g}$ electronic configuration. Stabilizing this in a square pyramidal environment would cost more energy than paying the penalty of correlation and delocalising the electrons. As pointed out earlier, getting the right structural distortion along with high correlation is needed for charge ordering, which does not seem to be possible for monolayer geometry. However, we do see charge ordering in electron doped bottom layers of thick films. We presume this is possible because of the octahedral geometry of the layer which makes it similar to bulk, to confirm this we therefore study electron doped bulk LVO.
		
	\subsection{Electron doped bulk \ce{LaVO3}}
\begin{figure}
		\centering
		\includegraphics[width=\linewidth]{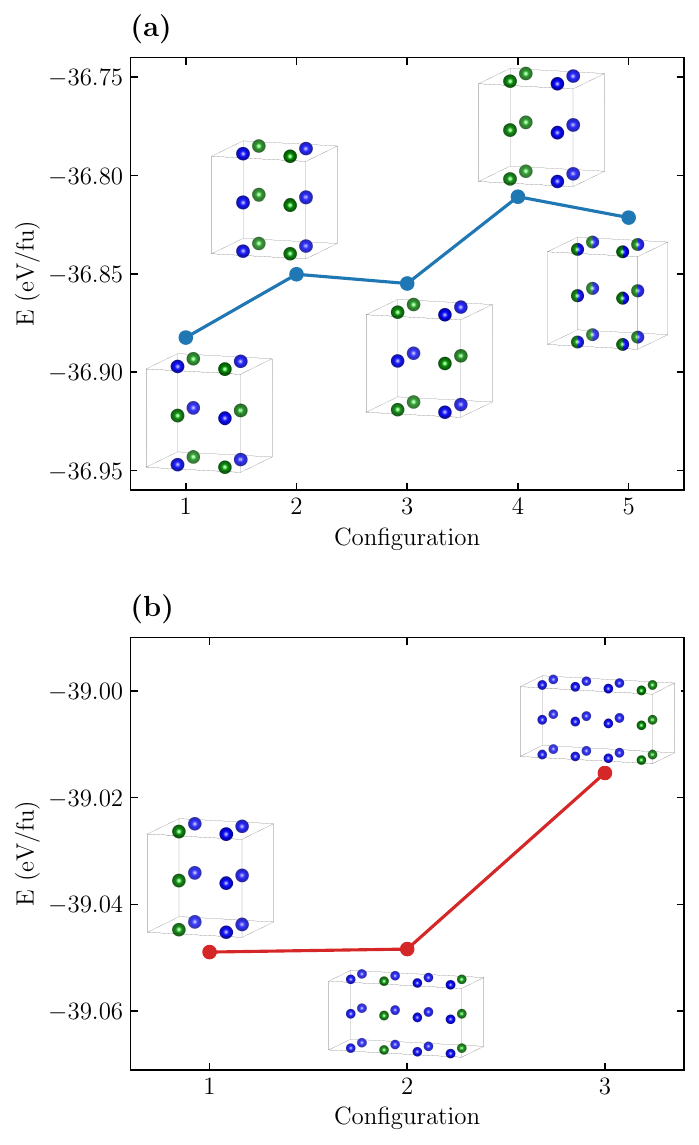}
		\caption{Relative energies of different charge ordering pattern considered at a doping concentration of (a) 0.5e/fu and (b) 0.25e/fu in the layered stripe-AFM state with ferromagnetic interlayer coupling. Blue sphere represents $\text{V}^{3+}$ and green are $\text{V}^{2+}$ }
		\label{fig6}
	\end{figure}

	We begin by introducing a doping concentration of $0.5e/\text{fu}$, which would yield a 1:1 ratio of $\text{V}^{3+}$ and $\text{V}^{2+}$. Because bulk LVO contains two vanadium layers per unit cell, four distinct charge-ordering patterns are possible. To determine the ground state, we compare the total energies of a uniform charge state against layered checkerboard and layered stripe arrangements, evaluating both in-phase and out-of-phase stackings for each. Compared to the monolayer, bulk LVO responds to electron doping by readily forming charge-ordered states. Across the various magnetic configurations considered, out-of-phase checkerboard charge ordering emerges as the ground state [Fig. \hyperref[fig6]{7(a)}]. This result confirms that the electron-doped bottom layers of thick films do exhibit charge ordering. However, we find no stripe-AFM layered arrangement in the ground state; instead, the C-AFM configuration remains lowest in energy even after electron doping.
	
	At a lower doping concentration of $0.25e/\text{fu}$, a 3:1 ratio of $\text{V}^{3+}$ and $\text{V}^{2+}$ is required. Consequently, we investigate the same ordering patterns discussed in the previous section. Although bulk LVO generally prefers out-of-phase stacking, multiple out-of-phase configurations are possible at this doping level. To focus specifically on planar charge-ordering patterns, we simplify the analysis by considering only in-phase stacking. Among the configurations tested, those with an ordering pattern analogous to the checkerboard arrangement emerge as the ground state with nearly degenerate energies [Fig. \hyperref[fig6]{7(b)}].
	
\section{Summary}
	In conlusion, we performed first-principles calculations to study magnetic and electronic structure property of freestanding \ce{LaVO3} films. Our methodology correctly reproduces the observed C-type AFM ground state with G-type orbital ordering for bulk \ce{LaVO3}. We find that the properties of ultrathin films strongly depend on the film thickness. In the monolayer, vanadium is five-fold coordinated, which removes the orbital ordering, presence of $d_{xy}$ orbital facilitates a next-nearest neighbor interaction which stabilizes a stripe-AFM state. In bilayer, due to the absence of $d_{xy}$ the top surface transitions to Neel-AFM, whereas since the bottom layer is octahedrally coordinated it retains the orbital ordering same as bulk and stays Neel-AFM. Since the top surface has no orbital ordering and bottom surface has orbital ordering the interlayer coupling is AFM. The trilayer behaves in the same way as bilayer, the only exception being that the interlayer coupling between the first two layer is AFM and bottom two layer is FM. Beyond four layer of \ce{LaVO3} a polar catastrophe driven charge transfer occurs that hole dopes the top surface which therefore transition to FM state and electron dopes the bottom surface which becomes stripe-AFM whereas the central layers behaves bulk like and have C-AFM configuration. We also observe that the charge transfer process is site selective hinting towards possible charge ordering. We confirm this by studying hole doped monolayer where we observe that at a doping concentration of 0.5h/fu a striped charge ordering pattern is preferred across all magnetic configuration considered and among them FM state had the lowest energy. Reducing the doping concentration to 0.25h/fu effectively increased the wavelength of charge ordering and we get a 3:1 striped charge ordered state. Electron doping the monolayer does not result in charge ordering, we assumed this is because of the five-fold coordination of monolayer that will not be able to stabilize V$^{2+}$ with a $t^3_{2g}$ electronic configuration. Therefore, we studied electron doped bulk \ce{LaVO3} and due to the octahedral coordination of vanadium in bulk structure it easily formed charge ordered state, confirming that the electron doped bottom layer of thick films does become charge ordered. At a doping concentration of 0.5e/fu, the checkerboard charge ordering with out-of-phase stacking came out to be the ground state. Reducing the concentration to 0.25e/fu resulted in a state analogous to checkerboard charge ordering.
	
	\ce{LaVO3} based heterostructures and superlattices serve as a prominent platform for observing exotic physical phenomena. However, a comprehensive theoretical understanding of these systems has remained limited. Our work on freestanding \ce{LaVO3} thin films provides valuable insights into these composite systems, offering a framework to design related materials and manipulate their functionalities. 
	
\section{Acknowledgement}
	We sincerely acknowledge the support and resources provided by the PARAM Brahma Facility at the Indian Institute of Science Education and Research, Pune, under the National Supercomputing Mission of the Government of India. A.A. acknowledges the Council of Scientific and Industrial Research (CSIR) India for support through a research fellowship.


%

\end{document}